\documentclass[twocolumn,epjc3]{svjour3}          

\RequirePackage[T1]{fontenc}
\smartqed  
\usepackage[utf8]{inputenc}
\RequirePackage{graphicx} 
\usepackage{subfig}
\usepackage{graphicx}
\usepackage{longtable}
\usepackage[abs]{overpic}
\usepackage{multirow}
\usepackage{braket}
\usepackage{lipsum}
\usepackage{comment}
\usepackage{amsmath}
\usepackage{amssymb}
\usepackage{siunitx}
\usepackage{booktabs}
\usepackage{upgreek}

\usepackage{color}   
\usepackage{hyperref}
\usepackage[switch]{lineno}
\usepackage{scalerel,amssymb}

\usepackage{ifthen}  
\newboolean{uprightparticles} 
\setboolean{uprightparticles}{false} 

\RequirePackage{upgreek}
\RequirePackage{graphicx}
\RequirePackage{mathptmx}      
\RequirePackage{flushend}
\RequirePackage[numbers,sort&compress]{natbib}
\usepackage[leftcaption]{sidecap}
\usepackage{graphics}
\pdfoutput=1
\usepackage{comment}
\usepackage{xcolor}
\usepackage{xurl}
\usepackage{float}
\usepackage{testhyphens}
\usepackage{hyphenat}
\usepackage{amsmath}
\usepackage{mathptmx}
\usepackage{orcidlink}
\journalname{Eur. Phys. J. C}

\makeatletter

\newcommand{\checknextarg}{\@ifnextchar\bgroup{\nolinebreak\gobblenextarg}{}}
\newcommand{\gobblenextarg}[1]{ \textsuperscript{\nolinebreak\hspace{-4pt}\mbox{\nolinebreak$^,$\nolinebreak\ref{#1}\nolinebreak}\nolinebreak} \@ifnextchar\bgroup{\gobblenextarg}{}}

\begin{document}

\title{SND@LHC Upgrade for the High-Luminosity LHC: Physics Reach and Installation Scenarios}

\author{The SND@LHC Collaboration\thanksref{e1}$^\text{\normalfont,1}$}
\thankstext{e1}{e-mail: Antonia.Di.Crescenzo@cern.ch}
\institute{Full authorlist supplied at the end of this article. \label{addr1}}




\maketitle 

\begin{abstract}
The SND@LHC experiment is currently taking data at the Large Hadron Collider (LHC), exploring the uni\-que forward region at pseudorapidities $7.2<\eta<8.4$. Its physics programme covers neutrinos originating from heavy-flavour decays and feebly interacting particles produced in proton proton collisions.  Building upon the successful operation of the present detector, this paper presents the physics reach of the approved SND@LHC upgrade for Run~4 of the LHC, and compares it with an alternative installation scenario.

Lowering the detector by approximately 40 cm and shifting it horizontally by about 30 cm, while keeping it off-axis, increases the total neutrino interaction rate  by a factor of five.  The paper describes  the design of the upgraded detector and compare the physics performance  in both installation scenarios.

\end{abstract}



\section{Introduction}

The LHC forward region has emerged as a unique environment to explore phenomena that are difficult or even impossible to study with the main detectors. The extremely boosted particles produced in proton–proton collisions at large pseudorapidities can travel hundreds of metres along the beamline, carrying valuable information on both Standard Model and beyond-Standard-Model (BSM) physics. 

Among these, high-energy neutrinos from heavy-flavour decay represent the most intriguing frontiers of the LHC physics programme.
The SND@LHC experiment (Scattering and Neutrino Detector at the LHC) was conceived to explore this forward window \cite{SNDTP,SNDLHC:2022ihg}. Installed in the TI18 tunnel, approximately 480 m downstream of the ATLAS interaction point, SND@LHC is designed to detect neutrinos produced in LHC collisions, predominantly from the decays of heavy-flavour hadrons. Since the start of data taking in 2022, the experiment has successfully demonstrated the feasibility of operating a compact hybrid detector in this radiation and background environment. The first observations of LHC neutrinos, both with and without a muon in the final state, have confirmed the scientific potential of this new experimental domain \cite{SNDLHC:2023pun,SNDLHC:2024qqb}.

The proposal for an upgrade of SND@LHC to operate throughout the High-Luminosity LHC \cite{Abbaneo:2926288} was approved by the Research Board in June 2025. 
The upgraded detector configuration and technologies are identical to the one proposed for the SND detector for SHiP and can be considered as a prototype for SHiP~\cite{Albanese:2948477}. The synergy between both experiments is thus particularly relevant.
For convenience, in the following the upgraded detector designed for operation at the HL-LHC is referred to as SND@HL-LHC. The name of the Collaboration remains unchanged, SND@LHC.
The HL-LHC will deliver an order of magnitude higher integrated luminosity, producing an unprecedented flux of forward neutrinos. 
This provides an extraordinary opportunity to perform precision measurements of neutrino interactions at energies up to a few TeV. 

To fully exploit this opportunity, an upgraded detector is required, capable of sustaining higher particle rates, improving background rejection, and providing enhanced kinematic reconstruction capabilities. The upgrade builds on the experience gained during Run 3, retaining the proven modular detector concept while optimizing its performance for the HL-LHC conditions.
In particular, the upgrade replaces the emulsion-based technology with fully electronic detectors to cope with the increased muon rates expected at the HL-LHC, and introduces a magnetic spectrometer to enable charge and momentum measurements, allowing neutrino and antineutrino interactions to be distinguished.

A central aspect of the proposal concerns the installation strategy within the existing TI18 tunnel. Two scenarios are compared, both based on the same upgraded detector design:
\begin{itemize}
\item The {\it Baseline configuration}, with the detector installed in the tunnel, without civil-engineering intervention. This is the baseline configuration for the approved version of the upgrade.
\item The  {\it Extended configuration}, 
in which the detector is lowered 40 cm and shifted horizontally by 27 cm to improve the detector positioning and acceptance.
\end{itemize}

The Baseline option represents the most straightforward and low-impact installation, requiring only mechanical adaptations to the available space. We will show that the Extended configuration
gives a substantial improvement in phy\-sics performance, with higher geometric acceptance and reduced background contamination.

The following sections describe the detector concept and technological design (Section~\ref{sec:detector}), the two installation scenarios (Section~\ref{sec:installation}), and the comparative performance studies (Section~\ref{sec:physics}). 
\section{Detector concept}
\label{sec:detector}

The upgraded detector maintains a total length of approximately 3 m, with a transverse cross-section of about 1 × 1 m², preserving the compact footprint that is essential for its integration in TI18. The detector axis is aligned with the forward direction from the ATLAS Interaction Point 1 (IP1), intercepting particles produced at pseudorapidities around $\eta \sim 7-8$. 

\begin{figure}[th]
    \centering
    \includegraphics[width=1.0\linewidth]{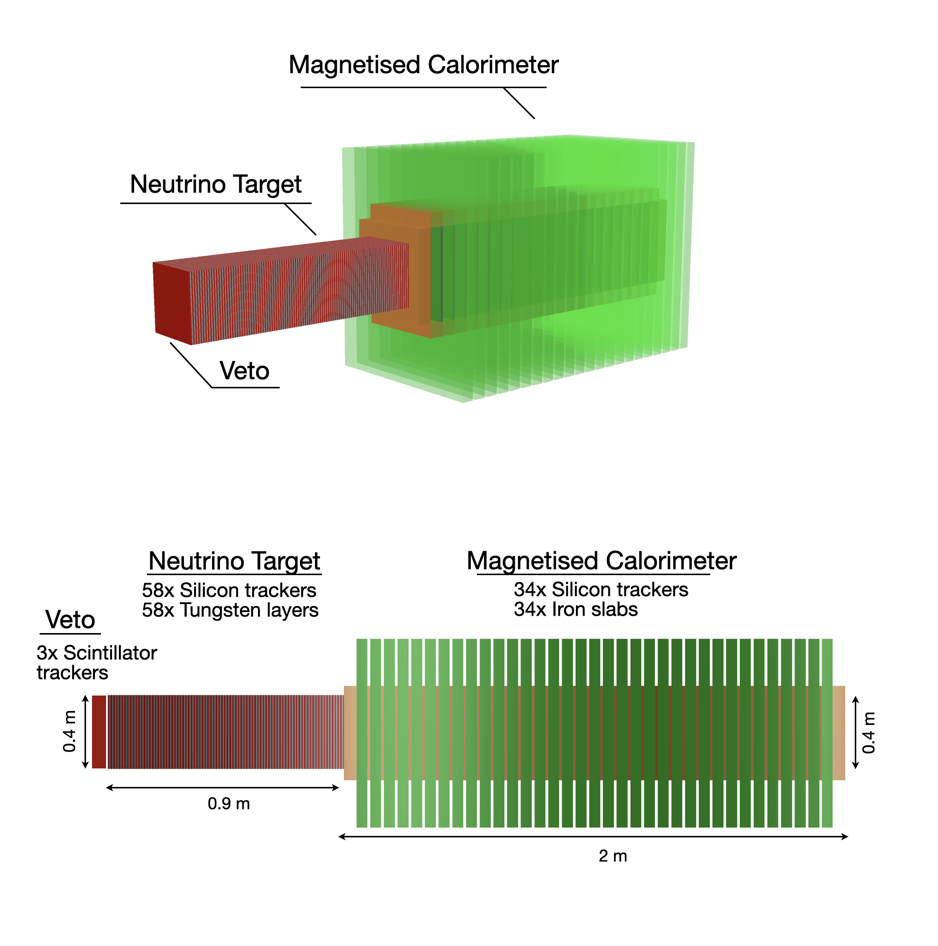}
    \caption{Layout of the SND@HL-LHC detector, as implemented in the 3D simulation: tiled view (top) and side view (bottom).}
    \label{fig:detector_layout}
\end{figure}

The SND@HL-LHC detector concept has been developed to efficiently identify all three neutrino flavours and measure their energy. To achieve these goals, the detector consists of a high-granularity and high-sampling calorimeter in the neutrino target region followed by a magnetised tracking calorimeter, as shown in Figure~\ref{fig:detector_layout}. The instrumented target region provides the measurement of the energy deposited therein and the reconstruction of the neutrino vertex with the necessary spatial resolution to efficiently contribute to the identification of the tau lepton. The magnetised tracking calorimeter is designed to measure the hadronic energy, identify muons produced in $\nu_\mu$ charged-current interactions and in the muonic decay of tau leptons, and measure their energy. Additional veto and fast-timing layers are located upstream and within the target region to tag charged backgrounds and allow high precision time correlation with the LHC bunch structure, respectively. 

\subsection{Veto system}

The veto system plays a crucial role in rejecting charged particles entering the detector acceptance, primarily high-energy muons originating from the ATLAS interaction point (IP1). 

SND@HL-LHC will employ the same basic veto concept successfully used during Run 3, based on fast plastic scintillator layers read out by silicon photomultipliers (SiPMs). The system is positioned upstream of the neutrino target and consists of three parallel planes of stacked scintillating bars, providing full coverage of the target’s 40 × 40 cm² active area. This configuration achieves nearly hermetic coverage and a rejection efficiency at the level of $10^{-9}$~\cite{SNDLHC:2025nrj}.


For the HL-LHC upgrade, the system will retain the same modular design and readout scheme, while benefiting from improved calibration and timing synchronisation with the LHC bunch structure.

\subsection{Timing system}
A dedicated timing system is integrated within the target region to provide fast response and precise time measurements, with a resolution better than 100 ps. This capability is essential to unambiguously associate detector activity with the correct LHC bunch crossing, suppressing accidental backgrounds from out-of-time particles and ensuring operation in a low-background, high-purity regime for neutrino detection and long-lived particle searches.

The timing system technology is currently under study, with two candidate solutions being investigated through dedicated R\&D activities, both capable of achieving the required time resolution; the final choice will be driven by performance and integration considerations.

\subsection{Neutrino target region}

The instrumentation of the neutrino target forms the core of the SND@HL-LHC detector. It is designed to provide both a dense medium for neutrino interactions and a finely segmented tracking calorimeter for track and vertex reconstruction, as well as for the measurement of the deposited energy. The target consists of a compact stack of 58 tungsten plates, each 7 mm thick, each followed by a silicon microstrip layer, covering an active area of nearly 40 × 40 cm². This structure corresponds to a total mass of about 1.3 tons, approximately 116 radiation lengths and 4 interaction lengths.

The tungsten plates serve as the neutrino interaction me\-dium, as passive material for the high-sampling calorimeter, and as mechanical supports for the silicon detector modules. The silicon modules are repurposed from the CMS Microstrip Tracker~\cite{CMS:1998aa}, with newly designed interconnection circuits adapted to the SND@HL-LHC geometry. The CMS strip modules provide spatial coordinates with a resolution better than 30 µm, enabling track and vertex reconstruction with sufficient precision to efficiently complement the kinematical event reconstruction for neutrino flavour identification, including tau neutrinos.

\begin{figure}[t]
    \centering
    \includegraphics[width=1.0\linewidth]{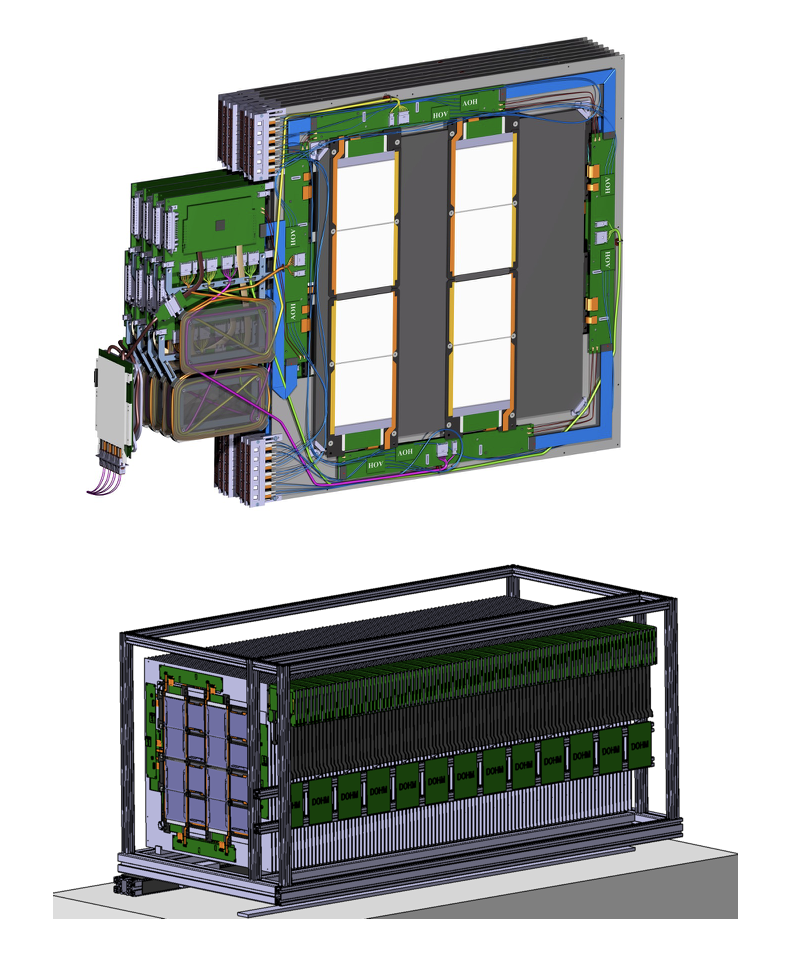}
    \caption{Top: a group of eight consecutive target stations connected to form a control ring. Bottom: 3D model of the neutrino target, including the mechanical support structure.}
    \label{fig:target_layout}
\end{figure}

Each detecting layer consists of eight silicon modules arranged in four quadrants, each with a side length of 186.1 mm, separated by inactive gaps 13.4 mm wide. This configuration results in an overall square active envelope with a side length of 385.6 mm and a fill factor of 93.2\%, defined as the fraction of the total envelope area covered by active silicon. The four quadrants are projective in all layers, ensuring that a neutrino interaction well contained within any quadrant is reconstructed with full efficiency in all subsequent target layers. Consecutive layers are rotated by 90 degrees, thus providing precision tracking in both spatial coordinates.
A group of eight consecutive stations connected through a single readout chain is shown in Fig.~\ref{fig:target_layout}, which also illustrates the supporting mechanical frame.

The electronics architecture closely follows that of the CMS Tracker. Dedicated interconnection circuits distribute power and control signals to the modules of a station, which are read out independently. Consecutive stations are grouped into a redundant control-ring architecture, with 8 to 10 stations per control ring.

\subsection{Magnetised calorimeter}

The magnetised hadronic calorimeter, located downstream of the neutrino target, plays a dual role in the SND@HL-LHC detector: it measures the energy of hadronic showers produced in neutrino interactions and provides charge and momentum measurements for muons.

The calorimeter follows the same basic detector concept as the neutrino target, with layers of absorber material interleaved with layers of silicon detector modules repurposed from the CMS Strip Tracker. In this region, however, the absorber consists of 50 mm thick iron slabs, chosen for their magnetic properties in order to achieve a sufficient integrated field. The detector, as implemented in the simulation, consists of 34 layers.

Given the significantly larger size and mass of the iron slabs compared to the tungsten plates of the target, the silicon modules are not mounted directly on the iron, but instead on separate steel plates with a thickness of 3 mm. This solution allows fully assembled and tested silicon layers to be inserted into the pre-assembled iron structure.




\begin{figure}[t]
    \centering
    \includegraphics[width=1.0\linewidth]{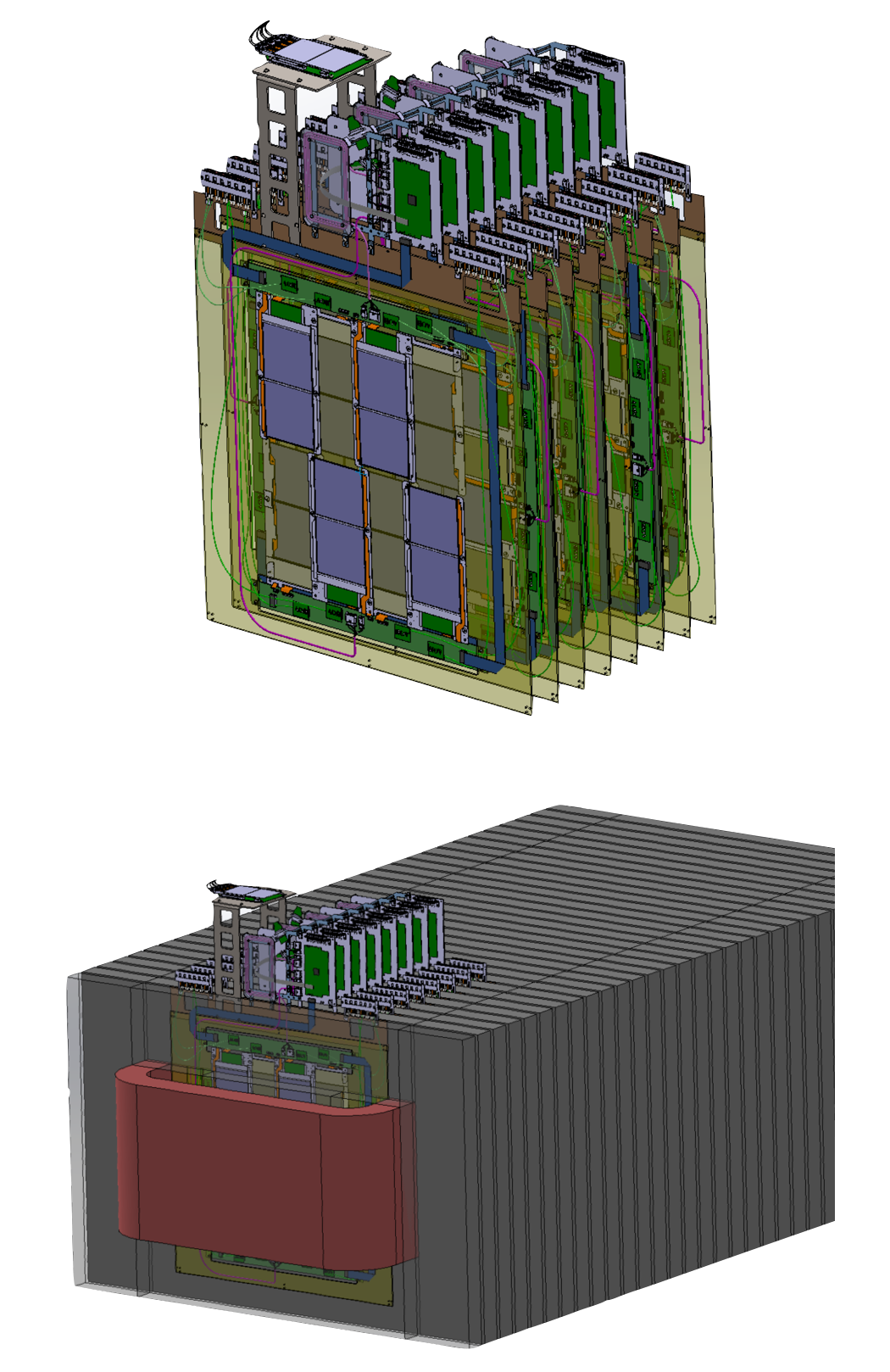}
    \caption{Top: a group of eight silicon detector layers; the first four layers are connected to form a control ring. Bottom: the silicon detector layers integrated with the iron structure and the coil.}
    \label{fig:calo_layout}
\end{figure}

The arrangement of the silicon modules and of the interconnection electronics closely follows that of the target, with two minor but significant differences introduced to optimise performance. First, geometrical inefficiencies in the calorimeter are made non-projective by slightly staggering consecutive layers. This ensures that at least three independent spatial coordinates are measured every four layers, guaranteeing full detection efficiency for through-going muons and a uniform response in the reconstruction of hadronic showers originating from neutrino interactions in the target. 
Second, in the upstream section of the calorimeter the orientation of the silicon strips alternates between horizontal and vertical, providing spatial sampling of hadronic showers in both views. In the downstream section, all microstrips are instead oriented parallel to the magnetic field direction in order to optimise the muon momentum resolution. A group of eight consecutive layers is shown in the top panel of Fig.~\ref{fig:calo_layout}.
.
The electronics architecture and readout system are identical to those of the target. In the calorimeter, four to six layers are connected to form a control ring, matching the larger longitudinal pitch of the layers. The use of identical silicon detectors and electronics systems for both the target and the iron calorimeter significantly simplifies detector operation and maintenance.

The calorimeter is embedded within a magnetised iron structure that provides a uniform magnetic field in the vertical direction. This field enables momentum measurements of muons with energies up to about 1 TeV, with a resolution of approximately 20\%, while keeping the overall detector length within the tight space constraints of the TI18 tunnel. Compactness was a key design driver: the coil and yoke geometry were optimised to minimise both the longitudinal footprint and the distance from the beam axis, while ensuring sufficient magnetic field strength for tracking. The use of a single-piece copper coil further simplifies transportation and assembly.

The iron structure with the coil and the integrated silicon detector layers is shown in the bottom panel of Figure~\ref{fig:calo_layout}. The system has been extensively simulated using three-dimensio\-nal finite-element models to optimise field uniformity and minimise stray fields in the surrounding environment. The results confirm a homogeneous magnetic field within the active region and a negligible external flux density, fully compatible with the TI18 operational environment.

The choice of magnet material balances performance, cost, and ease of manufacturing. Low-carbon steel provides the required field strength of approximately 1.7 T while keeping the power consumption at the kilowatt level. The power supply and cooling systems are based on standard CERN infrastructure, ensuring operational reliability and maintainability over the full HL-LHC running period.

Mechanically, the magnet system is segmented into sufficiently light components, each well below the lifting limits imposed by the TI18 access constraints. This modular segmentation allows the magnet to be transported in parts and assembled directly in the experimental area, while maintaining the required mechanical precision.

\section{Installation scenarios}
\label{sec:installation}

\begin{figure}[t]
    \centering
    \includegraphics[width=1.0\linewidth]{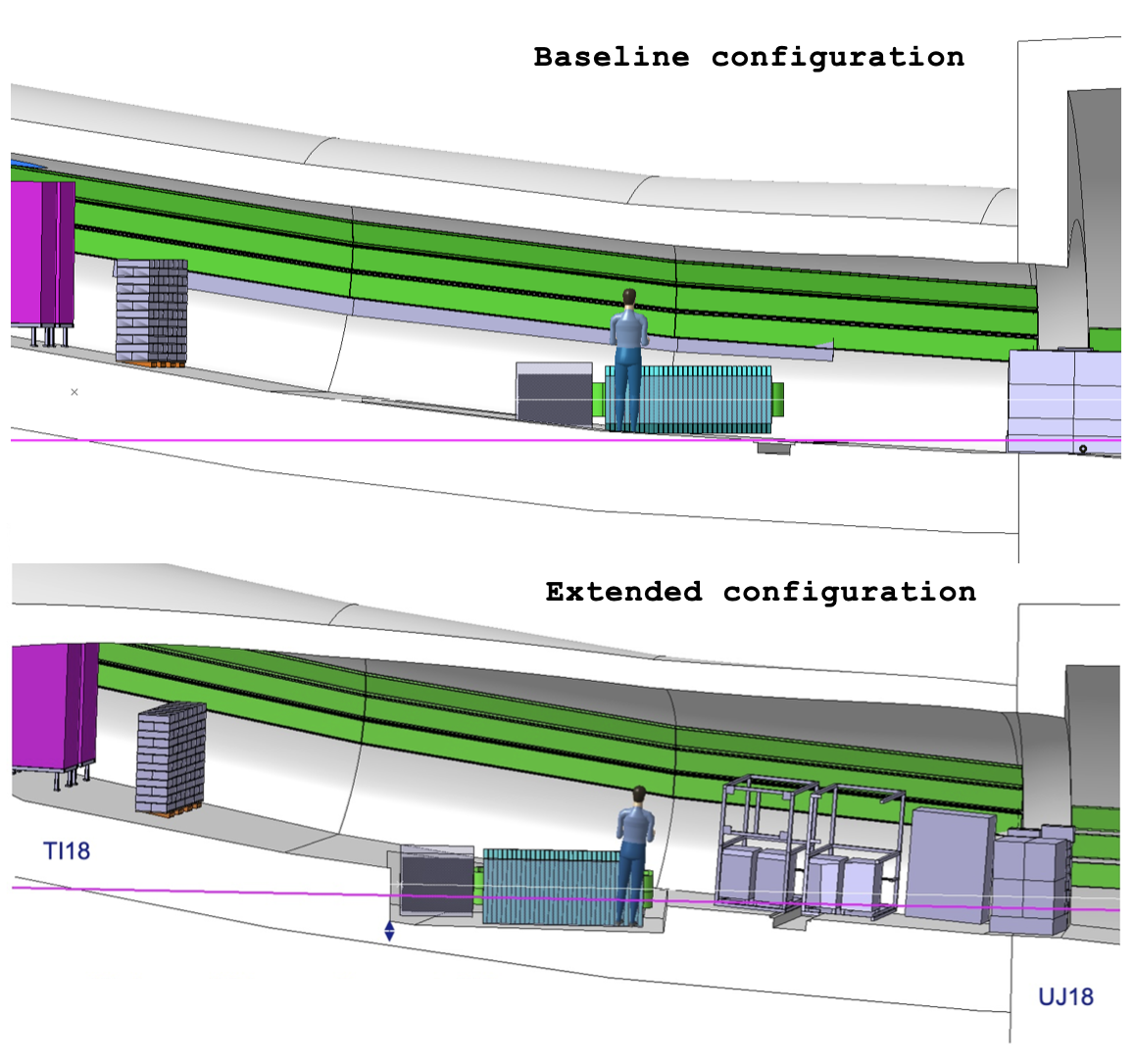}
    \caption{Integration of the SND@LHC upgrade in the Baseline configuration (top) and in the Extended configuration (bottom).}
    \label{fig:installation}
\end{figure}

Figure~\ref{fig:installation} shows the position of the SND@HL-LHC detector in the Baseline and Extended configurations. In both cases, the same detector design and readout systems are employed; the configurations differ only in the positioning of the detector within the TI18 tunnel.

In both layouts, the detector is intentionally placed off the collision axis. This configuration enhances the relative contribution of neutrinos from charm decays and provides complementarity with other forward experiments~\cite{FASER}. As a result, SND@HL-LHC probes a distinct region of phase space in terms of production mechanisms and neutrino energy spectra.

\subsection{The Baseline configuration}
The Baseline configuration represents the minimal-modifica\-tion option, installing the detector directly on the existing TI18 floor. In this layout, the detector sits slightly above the natural axis of the forward particle flux from IP1. Minor adjustments to the mechanical frame allow the alignment to be optimised within the available vertical clearance, while preserving compatibility with the tunnel’s existing infrastructure, ventilation, and access pathways.

\begin{figure}[h]
    \centering
    \includegraphics[width=0.7\linewidth]{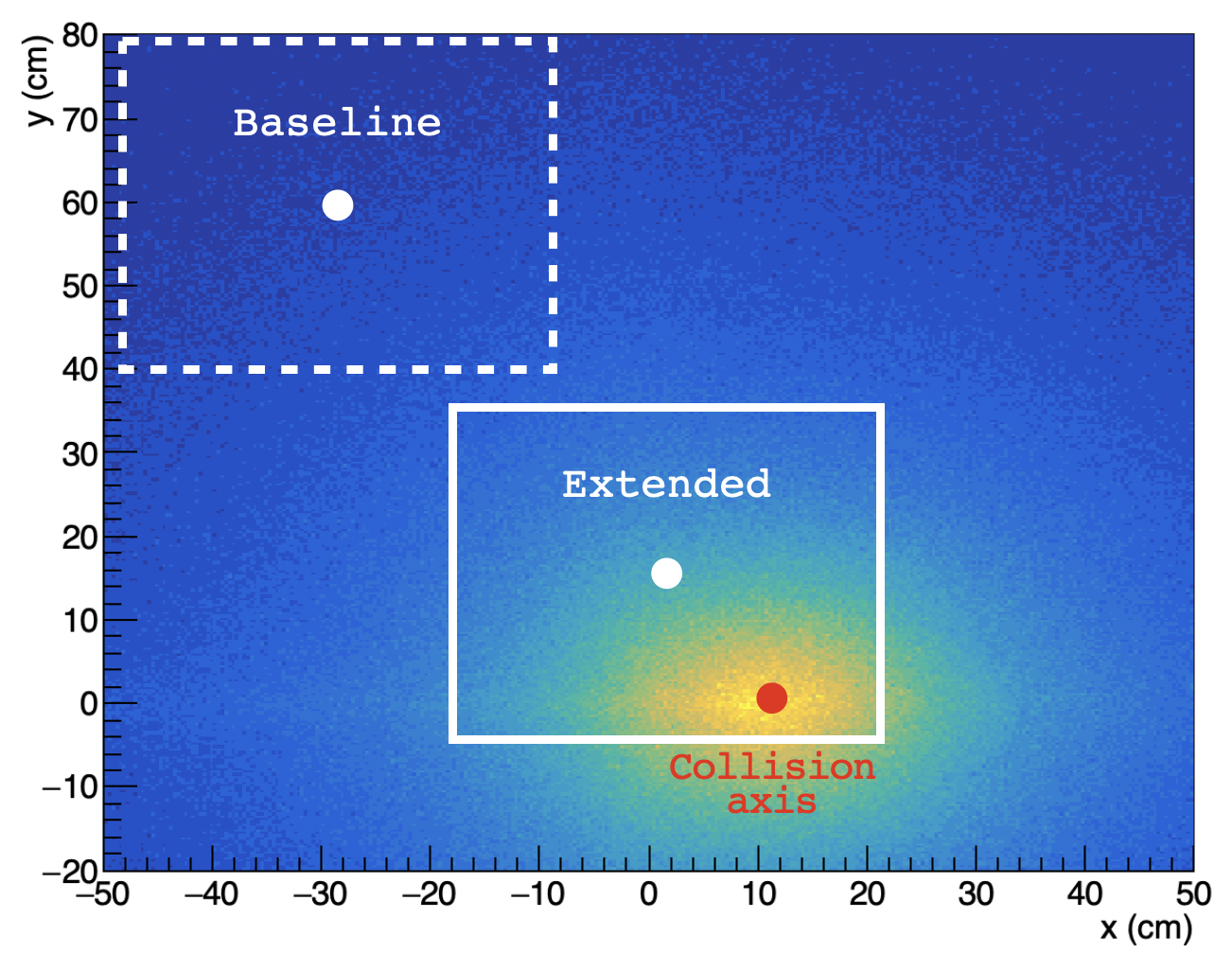}
    \caption{Position of the upgrade of the SND@LHC detector in the transverse plane superimposed to the $\nu_\mu$ flux predicted by simulations in HL-LHC for the Baseline (dashed line) and Extended (full line) configurations, assuming a +250 $\upmu$rad horizontal crossing angle.}
    \label{fig:acceptance}
\end{figure}

In the Baseline configuration, the detector centre is located at a transverse distance of approximately 57 cm in the vertical direction ($y$) and -39 cm in the horizontal direction ($x$) from the collision axis, corresponding to an average pseudorapidity coverage of $\eta$=[6.9,7.7], as shown in Figure~\ref{fig:acceptance}. The detector spans an azimuthal angle of $\Delta\phi \sim 66$ mrad, limited by the available vertical clearance in the tunnel.

Although this configuration avoids dedicated civil-engi\-neering work, it still requires adaptations to the existing infrastructure. In particular, the reduced vertical gap between the detector and the tunnel ceiling constrains the use of standard handling and lifting tools during installation and maintenance, requiring dedicated procedures and tooling solutions.

From a physics standpoint, the reduced azimuthal acceptance directly impacts the total neutrino interaction rate and the statistical reach at high energies, while preserving sensitivity to charm-induced neutrinos and maintaining complementarity with other forward experiments.


\subsection{The Extended configuration}

In the Extended configuration, 
the detector is lowered by  43 cm and shifted horizontally by  29 cm, while being kept off-axis. This adjustment improves the alignment with the axis of the forward particle flux.  In this configuration, 4.5 m$^3$ of concrete needs to be removed~\cite{lukasz,civil_talk}. . 


Lowering the detector in the Extended configuration reduces the transverse distance between the detector centre and the collision axis to 14 cm in the vertical direction ($y$) and 10 cm in the horizontal direction ($x$), while keeping the detector off-axis. This shift results in an extended pseudorapidity coverage ($\eta>7.8$) and a substantially larger azimuthal acceptance of $\Delta\phi \sim 250$ mrad. Compared to the Baseline configuration, the azimuthal coverage increases by roughly a factor of four, which constitutes the dominant contribution to the acceptance gain.



The combination of increased azimuthal coverage and reduced distance from the collision axis allows the detector to intercept a higher-energy neutrino component, resulting in a substantially higher interaction rate and enhanced sensitivity across the full HL-LHC programme, as discussed in detail in Section~\ref{sec:physics}.




\section{Physics performances}
\label{sec:physics}

The physics performance of the SND@HL-LHC detector has been evaluated through Monte Carlo simulations and parametric studies, building upon the validated simulation framework developed for Run 3~\cite{Roesler_2001,fluka,GENIE,Geant4}. These studies incorporate the expected beam and background conditions of the HL-LHC, including the full luminosity profile, particle fluxes in the TI18 tunnel, and detailed detector response of the proposed technological upgrades.

Figure~\ref{fig:acceptance} shows the transverse position of the upgraded detector superimposed on the simulated $\nu_\mu$ flux at the HL-LHC, for both the Baseline and Extended configurations, assuming a +250~$\upmu$rad horizontal crossing angle. 


\subsection{Neutrino detection}
The SND@HL-LHC detector will operate in an intense flux of high-energy neutrinos produced at the HL-LHC, providing a unique opportunity to explore neutrino interactions up to the TeV scale. The expected increase in integrated luminosity by about one order of magnitude compared to Run~3, together with the improved detector design, will translate into a substantial enhancement of the neutrino event yield and physics performances.

\begin{table}[b]
\centering
\begin{tabular}{c | c c | c c}
\toprule
 & \multicolumn{2}{c|}{\textbf{Baseline}} & \multicolumn{2}{c}{\textbf{Extended}} \\
\midrule
Flavour & Target & Target+HCAL & Target & Target+HCAL \\
\midrule
$\nu_\mu$         & 1.1$\times 10^4$ & 1.7$\times 10^4$ & 5.4$\times 10^4$  & 7.5$\times 10^4$  \\
$\bar{\nu}_\mu$   & 3.6$\times 10^3$ & 6.1$\times 10^3$ & 1.7$\times 10^4$  & 2.3$\times 10^4$  \\
$\nu_e$           & 1.7$\times 10^3$ & 2.8$\times 10^3$ & 7.3$\times 10^3$ & 9.7$\times 10^3$ \\
$\bar{\nu}_e$     & 7.6$\times 10^2$ & 1.2$\times 10^3$ & 2.7$\times 10^3$ & 4.4$\times 10^3$ \\
$\nu_\tau$        & 1.0$\times 10^2$ & 1.7$\times 10^2$ & 2.4$\times 10^2$ & 3.4$\times 10^2$ \\
$\bar{\nu}_\tau$  & 5.0$\times 10^1$ & 8.9$\times 10^1$ & 1.2$\times 10^2$ & 1.7$\times 10^2$ \\
\midrule
Total             & 1.7$\times 10^4$ & 2.7$\times 10^4$ & 8.1$\times 10^4$ & 1.1$\times 10^5$ \\
\bottomrule
\end{tabular}
\caption{
Predicted number of neutrino CC DIS interactions, as obtained from the \textsc{DPMJET} (light meson decays) and \textsc{POWHEG} (charmed hadron decays) simulations assuming 3000\,fb$^{-1}$ of integrated luminosity for the HL-LHC. Both the Baseline and Extended configurations are shown, distinguishing interactions in the target and in the combined target+HCAL volume.}
\label{tab:neutrino_yields}
\end{table}

\begin{figure}[t]
    \centering
    \includegraphics[width=1.0\linewidth]{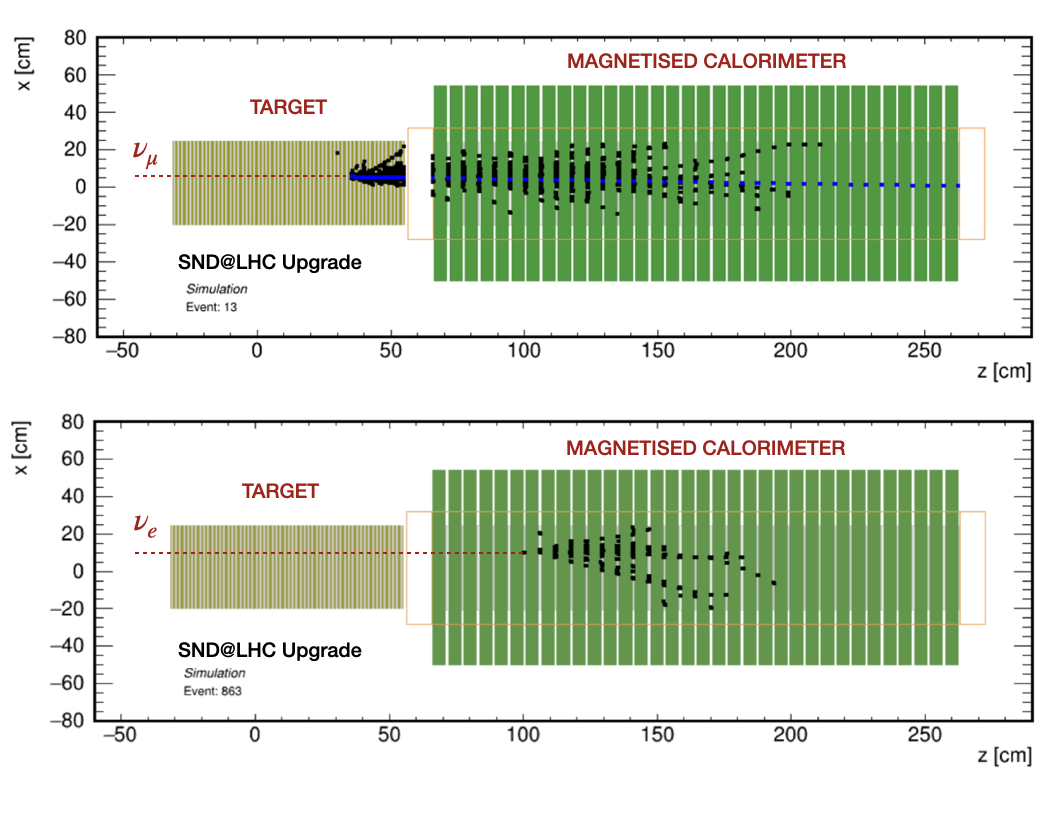}
    \caption{Displays of a simulated $\nu_\mu$ CC DIS interaction occurring in the neutrino target region (top) and a $\nu_e$ CC DIS interaction occurring in the magnetised calorimeter (bottom).}
    \label{fig:simulation}
\end{figure}

The expected neutrino yields and energy spectra for \linebreak SND@HL-LHC have been evaluated using the simulation framework developed for Run~3, based on the \linebreak \textsc{DPMJET}/\textsc{FLUKA} and \textsc{GENIE} generators, assuming an integrated luminosity of 3000 fb$^{-1}$ and a +250~\textmu rad horizontal beam configuration. The component originating from charmed hadron decays was predicted with the \linebreak \textsc{POWHEG+Pythia8} generators, restricted to hard QCD processes,

Event displays of neutrino interactions occurring in the target region and in the magnetised calorimeter are shown in Figure~\ref{fig:simulation}. The resulting interaction rates in the target and in the downstream calorimeter are summarised in Table~\ref{tab:neutrino_yields}, for both the Baseline and Extended configurations. Longitudinal shower containment is required for interactions in the HCAL.

In both cases, the detector will collect sizeable samples of neutrino interactions of all flavours, dominated by $\nu_\mu$ and $\nu_e$ produced in light-meson and charm decays. The HCAL contributes a significant additional sample of contained \linebreak events, enhancing the overall statistics.
Compared to the \linebreak Baseline configuration, the Extended layout increases the total neutrino interaction rate by approximately a factor of five, owing to its improved geometrical acceptance.

In addition to proton–proton collisions, heavy-ion runs at the HL-LHC will also produce forward neutrinos. Although the corresponding integrated luminosity is much \linebreak  smaller, the very high particle multiplicity in heavy-ion collisions could lead to a limited but potentially observable sample of neutrino interactions, which will be explored in future dedicated studies.

The effect of the LHC beam crossing angle on the neutrino flux has been studied. Three beam crossing angle configurations are foreseen for the HL-LHC operation: the nominal $+250$~\textmu rad horizontal configuration, which constitutes the baseline for Run~4, and the two vertical configurations with crossing angles of $\pm250$~\textmu rad. Among these, the \linebreak $+250$~\textmu rad vertical configuration maximises the neutrino  \linebreak yield, providing up to 60\% more CC~DIS interactions with respect to the nominal horizontal setting, owing to the reduced distance between the detector and the collision axis.

\begin{figure}[b]
    \centering
    \includegraphics[width=0.85\linewidth]{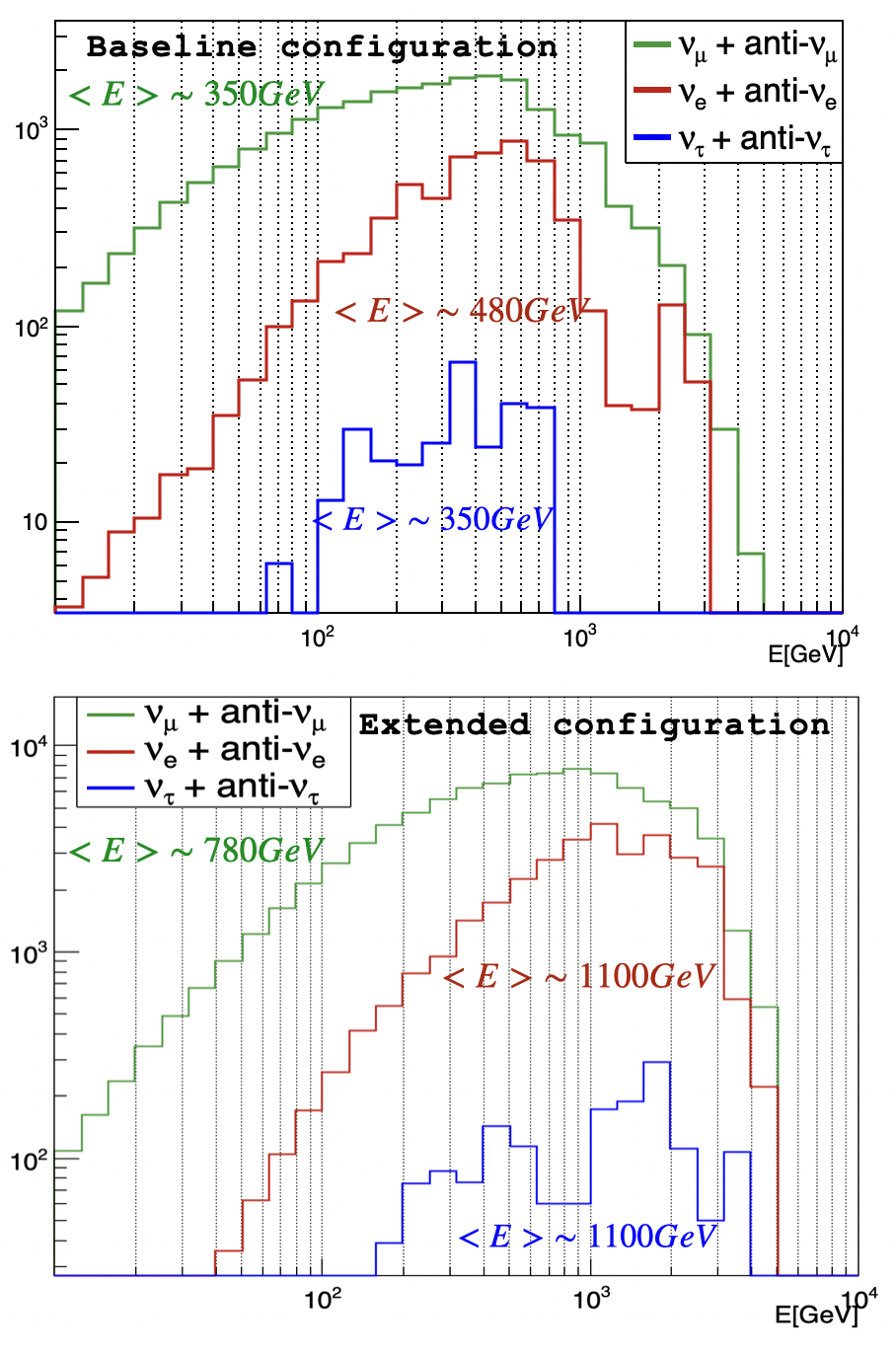}
    \caption{Energy spectra of the three neutrino flavours undergoing CC DIS interactions in the target for the Baseline (top) and Extended (bottom) configurations, assuming a +250$\upmu$rad horizontal crossing angle. The normalisation corresponds to 3000 fb$^{-1}$. Average energies are also reported.}
    \label{fig:spectra}
\end{figure}

Figure~\ref{fig:spectra} shows the simulated energy spectra of neutrinos interacting in the target. The comparison between the two configurations clearly illustrates the impact of the detector position relative to the beam axis. In the Extended configuration, the reduced distance to the collision axis leads to both a higher neutrino flux and a shift of the energy spectra toward larger values.
This effect directly translates into improved sensitivity to DIS processes and to the study of flavour composition at the highest accessible energies.

\subsection{Neutrino physics reach}

The expected precision on the main neutrino measurements achievable with the upgraded detector is summarised in Table~\ref{tab:neutrino_physics} for both the Baseline and Extended configurations.

\begin{table}[h!]
\centering
\begin{tabular}{l  c c c |  c c c}
\toprule
\multirow{2}{*}{Measurement}  & \multicolumn{3}{c|}{\textit{Baseline}} & \multicolumn{3}{c}{\textit{Extended}} \\
 & Stat. & Sys. & $\sigma$ & Stat. & Sys. & $\sigma$\\
\midrule
$\nu_\mu$, $\overline{\nu}_\mu$ cross-section & 3\% & 10\% & - & <1\% & 5\% & - \\                
Small-$x$ gluon PDF                           & 10\% & 5\% & - & 5\% & 5\% & -\\
Large-$x$ strange PDF                         & 6\%  & $<$1\% & - &3\% & $<$1\% & -\\
$\nu_e/\nu_\tau$  for LFU test                & 6\% & 5\% & - & 2\% & 5\% & -\\
$\nu_e/\nu_\mu$  for LFU test                 & 2\% & 5\% & - & 1\% & 2\% & -\\
$\overline{\nu}_\tau$ observation             & - & - & 3 & - & - & 5\\
\bottomrule
\end{tabular}
\caption{Expected precision for the main SND@LHC upgrade neutrino measurements in the Baseline and Extended configurations. Statistical and systematic uncertainties are reported.}
\label{tab:neutrino_physics}
\end{table}

\subsubsection{Muon neutrino and antineutrino cross-sections}
\begin{figure}[t]
    \centering
    \includegraphics[width=1.0\linewidth]{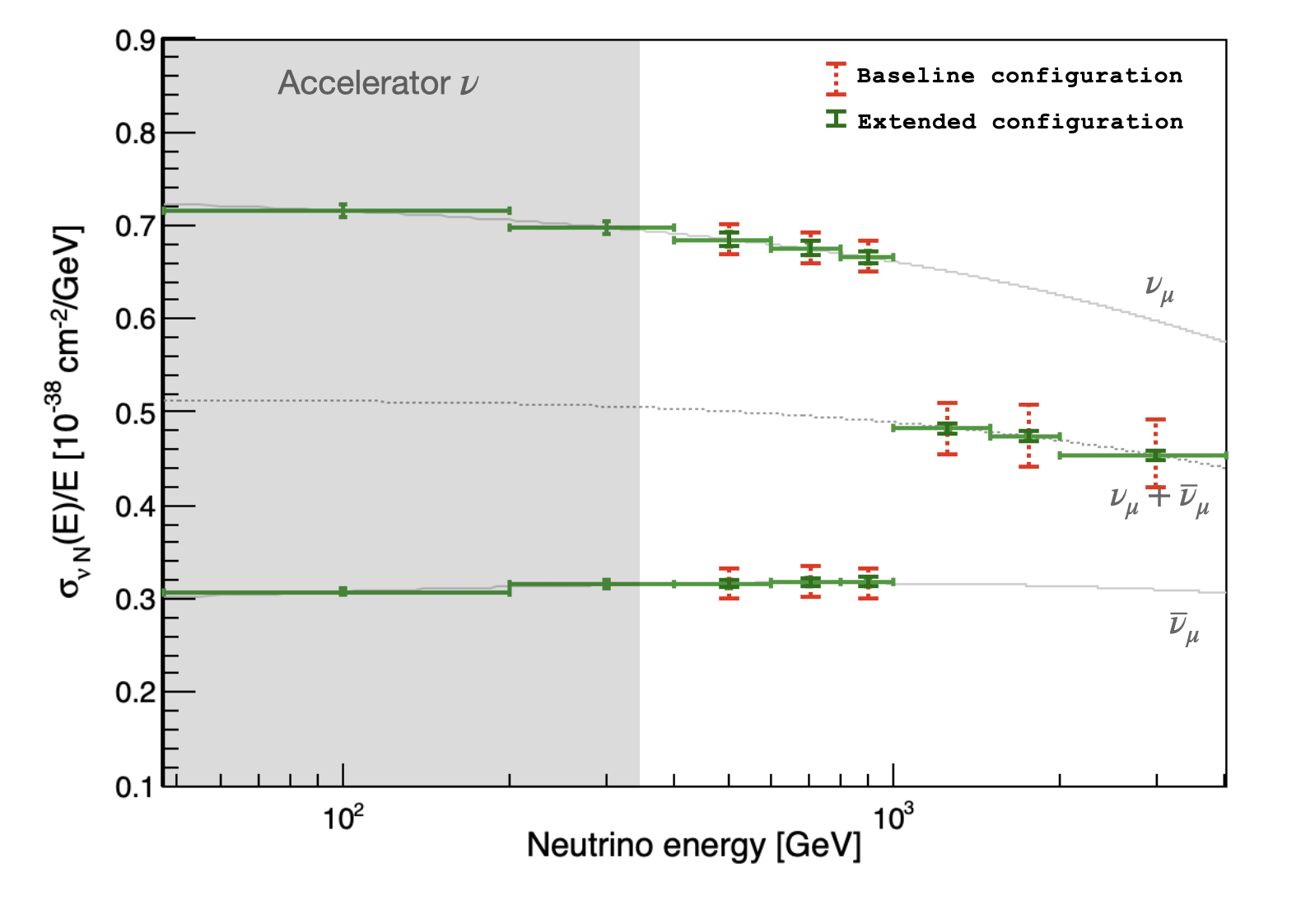}
    \caption{Expected statistical uncertainties for muon neutrino CC DIS cross-section measurements  in the Baseline and Extended configurations.}
    \label{fig:cross-section}
\end{figure}

SND@HL-LHC upgrade will measure the $\nu_\mu$ and $\bar{\nu}_\mu$ \linebreak CC~DIS cross-sections up to energies of the order of 1~TeV, thereby bridging the gap between fixed-target experiments and IceCube measurements (Figure~\ref{fig:cross-section}). The quoted statistical uncertainties refer to the previously unexplored energy region above 350~GeV. In the Baseline configuration, the statistical uncertainty in this range is expected to be around 3\%, while in the Extended configuration it improves to less than 1\%, benefiting from the larger geometric acceptance, higher neutrino yield, and extended sensitivity to higher neutrino energies.

The dominant systematic uncertainty arises from the \linebreak knowledge of the muon neutrino flux in the forward region. In the Extended configuration, this uncertainty can be Baseline to the level of about 5\%, thanks to the precise measurements of light meson production performed during Run\,3 by the LHCf Collaboration~\cite{Adriani:2012ap}, whose pseudorapidity coverage ($\eta \gtrsim 8.9$) largely overlaps with the acceptance of the Extended layout. In contrast, the Baseline configuration probes a more off-axis region, where the neutrino flux prediction depends more strongly on hadronic-interaction models. Comparisons among different models~\cite{Kling:2021gos} indicate a corresponding systematic uncertainty of approximately 10\% for the cross-section measurement.

The increased statistics of the Extended configuration also allow the cross-section measurement to be performed in a more differential manner, with a finer subdivision of the neutrino energy spectrum above 350~GeV. This opens the possibility to probe the energy dependence of the $\nu_\mu$ and $\bar{\nu}_\mu$ CC~DIS cross-sections with a granularity that would not be achievable in the Baseline configuration.

\subsubsection{Parton distribution functions}
Data collected by SND@HL-LHC will provide constraints on parton distribution functions (PDFs) via measurements of charm production in the forward regime of proton proton collisions, and also with measurements of neutrino deep inelastic scattering at TeV energies.

Electron neutrinos in the SND@HL-LHC pseudorapidity range originate predominantly from charm decays. Their measurement thus provides a sensitive probe of forward \linebreak charm production, a process affected by large uncertainties in the gluon parton distribution function (PDF) at very small Bjorken-$x$ ($<10^{-5}$). Improved constraints in this region are relevant both for future high-energy colliders such as the FCC and for predicting the flux of ultra-high-energy atmospheric neutrinos from charm decays~\cite{Bai:2022xad,Enberg:2016}.

Correlations between neutrino energy and pseudorapidity allow for PDF extractions largely insensitive to global normalisation uncertainties, such as those due to renormalisation and factorisation scales or the charm quark mass. The high statistics of the HL-LHC dataset will allow for shape measurements with a per-bin statistical uncertainty of the charm-related $\nu_e$ measurement around 10\% in the Baseline and 5\% in the Extended configuration, assuming 30 bins in the two-dimensional pseudorapidity and neutrino energy space. The residual per-bin systematic uncertainty is estimated to be about 5\%.

The measurement of neutrino-tungsten cross-sections at TeV energies provides constraints on proton PDFs at large values of Bjorken-$x$ ($>10^{-3}$)~\cite{Cruz-Martinez:2023sdv}. In particular, measurements of neutrino-induced charm production can significantly reduce the uncertainties on the strange quark PDF. The high granularity of the SND@HL-LHC silicon-instrumented target will allow for neutrino-induced charm production measurements via the identification of di-lepton final states $\mu\mu$, $e\mu$ and $ee$, where one of the leptons originates in the primary neutrino interaction and the other in the decay of a charm hadron, with a branching ratio of around 20\%. Around 240 neutrino-induced di-lepton events are expected in the Baseline configuration and 1145 in the enhanced configuration. To reduce the impact of systematic uncertainty to a negligible level, ratios of di-lepton to single-lepton cross-sections will be measured with expected statistical uncertainties \linebreak around 6\% for the Baseline and 3\% for the Extended configuration.

\subsubsection{Lepton flavour universality}
The identification of all three neutrino flavours enables precision tests of Lepton Flavour Universality (LFU). 

The fraction of electron neutrinos produced in pion and kaon decays is about 35\% (30\%) of the total number arriving at the target in the Baseline (Extended) configuration. Due to their lower energies and hence lower cross-sections, the number of neutrino interactions from neutrinos from pions and kaons is reduced to 20\% (13\%). Assuming that both tau and electron neutrinos come from the decay of charmed hadrons, the theory prediction of the $\nu_e$ to $\nu_\tau$ ratio ($R_{13}$) depends only on the decay branching ratios and the charm fractions, thus becoming sensitive to the cross-section ratio of the two neutrino species and allowing for a test of the lepton flavour universality in neutrino interactions. 
The ratio $R_{13}$ will be measured with a statistical uncertainty reduced from 30\% (Run 3) to about 6\% in the Baseline and 2\% in the Extended configuration.
Ongoing charm production studies~\cite{dstau,Akmete:2286844} are expected to bring the systematic uncertainty to the level of 5\%.

The $\nu_e/\nu_\mu$ ratio ($R_{12}$) can also probe LFU once low-energy neutrinos from pion and kaon decays are excluded. Using an energy threshold of about 600 GeV, the uncertainty on $R_{12}$ is expected to reach the 2\% level in the Baseline and 1\% in the Extended configurations. Further optimisation of the kinematic selection and external constraints from LHCf data~\cite{LHCf:2015rcj} will allow systematics to be reduced to 5\% and 2\%, respectively.

\subsubsection{Coincident detection of neutrinos with ATLAS events}
The large neutrino sample expected at SND@HL-LHC also enables the identification of neutrino interactions in coincidence with ATLAS events. In about 10\% of neutrino interactions originating in open charm production, simulations show that associated particles are produced within the ATLAS detector acceptance. If identified in ATLAS, these events provide exceptionally clean flavour samples dominated by heavy-flavour decays, allowing LFU tests with reduced systematic uncertainty. 

The feasibility of this measurement relies on anticipated improvements in forward tracking and vertexing expected over the next decade in the context of the HL-LHC upgrade programme. Timing synchronisation between \linebreak SND@HL-LHC and the ATLAS HGTD, with resolutions below 50~ps, will be essential for matching such events. With an expected yield of about 600 coincidences in the Baseline and 1300 in the Extended configuration over the HL-LHC run, these samples of coincident events will directly link forward measurements in general purpose collider detectors to neutrino detection in SND@LHC.

\subsubsection{Tau antineutrino observation}

The large neutrino flux available at the HL-LHC provides, for the first time, the opportunity to observe tau antineutrinos, which have not yet been directly observed experimentally. In the Baseline configuration, about 50 $\bar{\nu}_\tau$ charged-current interactions are expected in the neutrino target region over the full HL-LHC dataset. Restricting the analysis to the muonic decay channel, where the discrimination between neutrinos and antineutrinos can be achieved directly through the measurement of the charge of the primary lepton, reduces the sample to approximately 8 events, given the corresponding branching ratio. The application of a preliminary flavour-identification strategy based on multivariate techniques, developed in \cite{SNDTP}, further reduces the expected signal to about 4 observable events. The dominant background arises from misidentified $\bar{\nu}_\mu$ charged-current interactions and is estimated to be approximately 0.4 events, corresponding to a misidentification probability of about $10^{-4}$. As a result, the statistical significance of a $\bar{\nu}_\tau$ observation in this configuration is of the order the $3\sigma$ level.

On the contrary, the Extended configuration significantly improves the sensitivity to tau antineutrinos. In this case, approximately 120 $\bar{\nu}_\tau$ charged-current interactions are expected in the neutrino target region, leading to about 9 observable signal events after applying the same selection criteria. The corresponding background is estimated to be \linebreak around 2 events, yielding an observation significance of about $5\sigma$. 

Beyond the muonic decay channel, additional information from kinematic variables can be used, such as the inelasticity $y$, defined as the fraction of the neutrino energy transferred to the hadronic system. The different $y$ distributions of neutrino and antineutrino charged-current interactions provide statistical discrimination even in channels where the lepton charge is not measured. This approach could be exploited in high-statistics samples to constrain the different components through global fits. This consideration further motivates the Extended configuration, for which the substantially larger event sample would enable more differential and robust analyses. In this paper, however, a conservative strategy based solely on the $\tau \to \mu$ channel is adopted.



\subsection{Long-lived and feebly interacting particle searches}

The SND@HL-LHC detector also provides sensitivity to new long-lived or feebly interacting particles (FIPs) produced in the far-forward region of the LHC. Well-motivated scenarios include dark photons ($V$), heavy neutral leptons (HNLs), and axion-like particles (ALPs), which can be produced in proton--proton collisions at the LHC interaction point, propagate over long distances, and either decay or scatter inside the detector volume.

\begin{figure}[b]
    \centering
    \includegraphics[width=1.0\linewidth]{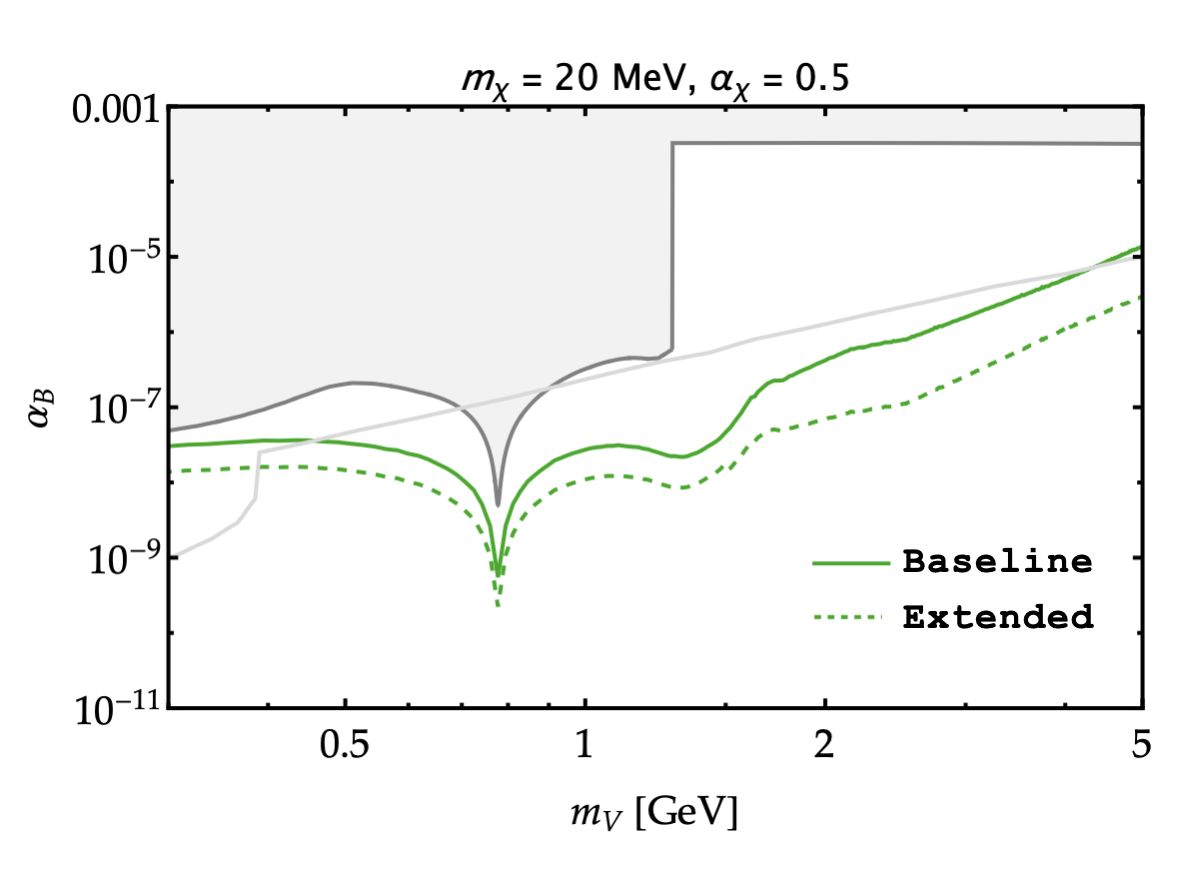}
    \caption{$2\sigma$ sensitivity of SND@HL-LHC to light dark matter models coupled to a baryonic mediator, for inelastic scattering off protons in the Baseline and Extended configurations. The grey region represents the parameter space excluded by other experiments~\cite{Boyarsky:2021moj}.}
    \label{fig:LDM}
\end{figure}

As a representative benchmark, the sensitivity of \linebreak SND@HL-LHC to light dark matter (LDM) particles $\chi$ interacting via a leptophobic $U(1)_B$ mediator is evaluated for both the Baseline and Extended configurations. The analysis considers inelastic scattering of LDM off protons in the detector material, $\chi + p \rightarrow \chi' + X$, which leads to neutral-current--like signatures and an excess of neutrino-like scattering events over the Standard Model expectation. The resulting $2\sigma$ sensitivity for $m_\chi=20$ MeV and $\alpha_\chi=0.5$ is shown in Figure~\ref{fig:LDM}.

Compared to the Baseline configuration, the Extended layout benefits from the increased geometrical acceptance and reduced distance from the collision axis, resulting in a higher LDM flux intersecting the detector. This leads to a significant improvement in sensitivity to the mediator coupling, extending the reach of SND@HL-LHC in searches for feebly interacting particles in the far-forward region.

\section{Conclusions and outlook}



The SND@HL-LHC detector builds on the experience gai\-ned during Run~3, retaining the same modular layout and detection principles, while incorporating several key improvements. These include an increased active target mass, fully electronic readout systems capable of operating at the higher muon rates of the HL-LHC, enhanced tracking and timing performance, and the addition of a magnetic spectrometer enabling charge and momentum measurements and the separation of neutrino and antineutrino interactions.

Two installation scenarios have been explored in detail: the Baseline configuration, which fits the detector within the existing TI18 tunnel geometry, and the Extended configuration, which  allows an increased geometric acceptance, resulting in an increase of the total neutrino interaction rate by approximately a factor of five.
Both configurations satisfy the fundamental integration and safety requirements of the LHC infrastructure. However, detailed simulation studies show that the Extended configuration provides a significant improvement in physics reach, with higher acceptance for forward neutrinos and enhanced sensitivity to feebly interacting particles.


The development and operation of the SND@HL-LHC detector will provide valuable experience in detector technologies, integration and data analysis that is directly relevant for the SND detector at SHiP. This synergy, which is independent of the specific installation configuration discussed in this paper, applies to both the Baseline and Extended layouts.
\section*{Acknowledgments}

We acknowledge the support for the construction and operation of the SND@LHC detector provided by the following funding agencies:  CERN;  the Bulgarian Ministry of Education and Science within the National
Roadmap for Research Infrastructures 2020–2027 (object CERN); ANID FONDECYT grants No. 3230806, No. 1240066, 1240216 and ANID  - Millenium Science Initiative Program -  $\rm{ICN}2019\_044$ \linebreak
(Chile); the Deutsche Forschungsgemeinschaft (DFG, ID \linebreak
496466340); the Italian National Institute for Nuclear Physics (INFN); JSPS, MEXT, the~Global COE program of Nagoya University, the~Promotion
and Mutual Aid Corporation for Private Schools of Japan for Japan;
the National Research Foundation of Korea with grant numbers 
\linebreak
2021R1A2C2011003, 2020R1A2C1099546,
\linebreak
2021R1F1A1061717, and 2022R1A2C100505; Fundação
\linebreak
para a Ciência e a Tecnologia, FCT grant numbers 
\linebreak
CEECIND/01334/2018, 
CEECINST/00032/2021 and
\linebreak
PRT/BD/153351/2021 (Portugal), 
CERN/FIS-INS/0028/2021; the Swiss National Science Foundation (SNSF); TENMAK for Turkey (Grant No.
2022TENMAK(CERN) A5.H3.F2-1).
J.C.~Helo~Herrera and O.~J.~Soto~Sandoval  acknowledge support from ANID 
FONDECYT grants No.1241685 and 1241803.
M.~Climesu, H.~Lacker and R.~Wanke are funded by the Deutsche 
Forschungsgemeinschaft (DFG, German Research Foundation), Project 496466340. This research was financially supported by the Italian Ministry of University and Research within the Prin 2022 program, Grant Assignment Decree No. 974 adopted on 30/06/2023.

We express our gratitude to our colleagues in the CERN accelerator departments for the excellent performance of the LHC. We thank the technical and administrative staff at 
\linebreak 
CERN and at other SND@LHC institutes for their contributions to the success of the SND@LHC efforts. We thank Luis Lopes, Jakob Paul Schmidt and Maik Daniels for their help during the~construction.

\bibliographystyle{epjc}
\bibliography{references}

@inproceedings{Roesler_2001,
      author         = "Roesler, Stefan and Engel, Ralph and Ranft, Johannes",
      title          = "{The Monte Carlo event generator DPMJET-III}",
      booktitle      = "{Advanced Monte Carlo for radiation physics, particle
                        transport simulation and applications. Proceedings,
                        Conference, MC2000, Lisbon, Portugal, October 23-26,
                        2000}",
      url            = "http://www-public.slac.stanford.edu/sciDoc/ docMeta.aspx?slacPubNumber=SLAC-PUB-8740",
      year           = "2000",
      pages          = "1033",
      doi            = "10.1007/978-3-642-18211-2_166",
      eprint         = "hep-ph/0012252",
      archivePrefix  = "arXiv",
      primaryClass   = "hep-ph",
      reportNumber   = "SLAC-PUB-8740",
      SLACcitation   = "%%CITATION = HEP-PH/0012252;%%"
}

@Article{ SNDTP,
 author = "Ahdida, C. and others",
 title =" {TECHNICAL PROPOSAL, SND@LHC, Scattering and Neutrino Detector at the LHC}",
 journal = "{ CERN LHC-2021-003, LHCC-P-016}",
 year    = "{24 Feb. 2021}",

 }

@article{SNDLHC:2024qqb,
    author = "Abbaneo, D. and others",
    collaboration = "SND@LHC",
    title = "{Observation of Collider Neutrinos without Final State Muons with the SND@LHC Experiment}",
    eprint = "2411.18787",
    archivePrefix = "arXiv",
    primaryClass = "hep-ex",
    reportNumber = "CERN-EP-2024-316",
    doi = "10.1103/r2qy-9hft",
    journal = "Phys. Rev. Lett.",
    volume = "134",
    number = "23",
    pages = "231802",
    year = "2025"
}

@article{GENIE,
    author = "Andreopoulos, C. and others",
    title = "{The GENIE Neutrino Monte Carlo Generator}",
    eprint = "0905.2517",
    archivePrefix = "arXiv",
    primaryClass = "hep-ph",
    reportNumber = "FERMILAB-PUB-09-418-CD",
    doi = "10.1016/j.nima.2009.12.009",
    journal = "Nucl. Instrum. Meth. A",
    volume = "614",
    pages = "87--104",
    year = "2010"
}

@article{Geant4,
      author         = "Agostinelli, S. and others",
      title          = "{GEANT4: A Simulation toolkit}",
      collaboration  = "GEANT4",
      journal        = "Nucl. Instrum. Meth. A",
      volume         = "506",
      year           = "2003",
      pages          = "250",
      doi            = "10.1016/S0168-9002(03)01368-8",
      SLACcitation   = "%%CITATION = NUIMA,A506,250;%%"
}

@techreport{fluka,
      author         = "Ferrari, Alfredo and Sala, Paola R. and Fasso, Alberto and Ranft, Johannes",
      title          = "{FLUKA: A multi-particle transport code (Program version 2005)}",
      institution    = "CERN",
      address       = "Geneva",
      doi            = "10.2172/877507",
      year           = "2005",
      number   = "CERN-2005-010, SLAC-R-773, INFN-TC-05-11",
      SLACcitation   = "%%CITATION = CERN-2005-010;%%"
}

@Article{ FASER,
 author = "Ariga, A. and others",
 title =" {TECHNICAL PROPOSAL, FASER, FORWARD SEARCH EXPERIMENT AT THE LHC}",
 journal = "{ CERN LHC-2018-036, LHCC-P-013}",
 year    = "{7 Nov. 2018}",
 eprint         = "1812.09139v1",
 archivePrefix  = "arXiv"
 }

@Article{ DPMJET,
 author = " Fedynitch, A.",
 title   = "{Cascade equations and hadronic interactions at very high energies}",
 journal = " CERN-THESIS-2015-371 {\url{https://cds.cern.ch/record/2231593}}",
 year = "27/11/2015"
}

@article{SNDLHC:2023pun,
    author = "Albanese, R. and others",
    collaboration = "SND@LHC",
    title = "{Observation of Collider Muon Neutrinos with the SND@LHC Experiment}",
    eprint = "2305.09383",
    archivePrefix = "arXiv",
    primaryClass = "hep-ex",
    reportNumber = "CERN-EP-2023-092",
    doi = "10.1103/PhysRevLett.131.031802",
    journal = "Phys. Rev. Lett.",
    volume = "131",
    number = "3",
    pages = "031802",
    year = "2023"
}

@article{Adriani:2012ap,
    author = "Adriani, O. and others",
    collaboration = "LHCf",
    title = "{Measurement of forward neutral pion transverse momentum spectra for $\sqrt{s}$ = 7TeV proton-proton collisions at LHC}",
    eprint = "1205.4578",
    archivePrefix = "arXiv",
    primaryClass = "hep-ex",
    doi = "10.1103/PhysRevD.86.092001",
    journal = "Phys. Rev. D",
    volume = "86",
    pages = "092001",
    year = "2012"
}

@article{dstau,
    author = {Aoki, Shigeki and others},
    year = {2020},
    month = {01},
    pages = {},
    title = {DsTau: study of tau neutrino production with 400 GeV protons from the CERN-SPS},
    volume = {2020},
    journal = {Journal of High Energy Physics},
    doi = {10.1007/JHEP01(2020)033}
}

@article{LHCf:2015rcj,
  title = {Measurements of longitudinal and transverse momentum distributions for neutral pions in the forward-rapidity region with the LHCf detector},
  author = {Adriani, O. and Berti, E. and Bonechi, L. and Bongi, M. and D'Alessandro, R. and Del Prete, M. and Haguenauer, M. and Itow, Y. and Iwata, T. and Kasahara, K. and Kawade, K. and Makino, Y. and Masuda, K. and Matsubayashi, E. and Menjo, H. and Mitsuka, G. and Muraki, Y. and Papini, P. and Perrot, A.-L. and Ricciarini, S. and Sako, T. and Sakurai, N. and Suzuki, T. and Tamura, T. and Tiberio, A. and Torii, S. and Tricomi, A. and Turner, W. C. and Ueno, M. and Zhou, Q. D.},
  collaboration = {LHCf Collaboration},
  journal = {Phys. Rev. D},
  volume = {94},
  issue = {3},
  pages = {032007},
  numpages = {38},
  year = {2016},
  month = {Aug},
  publisher = {American Physical Society},
  doi = {10.1103/PhysRevD.94.032007},
  url = {https://link.aps.org/doi/10.1103/PhysRevD.94.032007}
}

@article{SNDLHC:2025nrj,
    author = "Abbaneo, D. and others",
    collaboration = "SND@LHC",
    title = "{Installation and performance of the 3rd Veto plane at the SND@LHC detector}",
    eprint = "2502.10188",
    archivePrefix = "arXiv",
    primaryClass = "physics.ins-det",
    doi = "10.1088/1748-0221/20/07/P07011",
    journal = "JINST",
    volume = "20",
    number = "07",
    pages = "P07011",
    year = "2025"
}

@article{Bai:2022xad,
    author = "Bai, Weidong and Diwan, Milind and Garzelli, Maria Vittoria and Jeong, Yu Seon and Kumar, Karan and Reno, Mary Hall",
    title = "{Forward production of prompt neutrinos from charm in the atmosphere and at high energy colliders}",
    eprint = "2212.07865",
    archivePrefix = "arXiv",
    primaryClass = "hep-ph",
    reportNumber = "BNL-224879-2023-JAAM",
    doi = "10.1007/JHEP10(2023)142",
    journal = "JHEP",
    volume = "10",
    pages = "142",
    year = "2023"
}

@article{Boyarsky:2021moj,
    author = "Boyarsky, Alexey and Mikulenko, Oleksii and Ovchynnikov, Maksym and Shchutska, Lesya",
    title = "{Searches for new physics at SND@LHC}",
    eprint = "2104.09688",
    archivePrefix = "arXiv",
    primaryClass = "hep-ph",
    doi = "10.1007/JHEP03(2022)006",
    journal = "JHEP",
    volume = "03",
    pages = "006",
    year = "2022"
}

@article{Enberg:2016,
    author = "Ennerg,R et al",
    title = "{Prompt atmospheric neutrino fluxes: perturbative QCD models and nuclear effects}",
    journal = "JHEP",
    volume = "11",
    pages = "167",
    year = "2016"
}

@techreport{Abbaneo:2926288,
      author        = "Abbaneo, D and others",
      collaboration = "SND@LHC",
      title         = "{SND@HL-LHC, Scattering and Neutrino Detector in Run 4 of the LHC}",
      institution   = "CERN",
      number  = "CERN-LHCC-2025-004, LHCC-P-026",
      address       = "Geneva",
      year          = "2025",
      url           = "https://cds.cern.ch/record/2926288",
}

@article{SNDLHC:2022ihg,
    author = "Acampora, G. and others",
    collaboration = "SND@LHC",
    title = "{SND@LHC: the scattering and neutrino detector at the LHC}",
    eprint = "2210.02784",
    archivePrefix = "arXiv",
    primaryClass = "hep-ex",
    doi = "10.1088/1748-0221/19/05/P05067",
    journal = "JINST",
    volume = "19",
    number = "05",
    pages = "P05067",
    year = "2024"
}

@techreport{lukasz,
      author         = "{L. Krzempek}",
      title          = "{Integration study of Advanced SND experiment in TI18}",
      institution   = "CERN",
      month           = "07 June",
      year           = "2024",
      url       ={https://edms.cern.ch/ui/file/3076501},
      type          = "EDMS",
      number         = "LHC-X1SND-EN-0003",
}

@article{Kling:2021gos,
    author = "Kling, Felix and Nevay, Laurence J.",
    title = "{Forward neutrino fluxes at the LHC}",
    eprint = "2105.08270",
    archivePrefix = "arXiv",
    primaryClass = "hep-ph",
    doi = "10.1103/PhysRevD.104.113008",
    journal = "Phys. Rev. D",
    volume = "104",
    number = "11",
    pages = "113008",
    year = "2021"
}

@techreport{CMS:1998aa,
    collaboration = "CMS",
    title = "{CMS, tracker technical design report}",
 institution   = "CERN",
    number = "CERN-LHCC-98-06, CMS-TDR-5",
    year = "1998"
}

@techreport{Akmete:2286844,
      collaboration        = "{SHiP}",
      title         = "{Measurement of associated charm production induced by 400 GeV/c protons}",
      number        = "CERN-SPSC-2017-033, SPSC-EOI-017",
      institution   = "CERN",
      year          = "2017",
      type		= "Scientific Committee Paper",
      reportNumber  = "CERN-SPSC-2017-033",
      url           = "http://cds.cern.ch/record/2286844",
      address           = "\url{http://cds.cern.ch/record/2286844}",
}

@misc{civil_talk,
  author       = {T. Bud and J. Osborne},
  title        = {Civil Engineering aspects of the SND@LHC upgrade},
  howpublished = {Talk presented at the SND@LHC Collaboration Meeting},
  organization = {CERN},
  address      = {Geneva, Switzerland},
  month        = {Dec},
  year         = {2024}}

@article{Cruz-Martinez:2023sdv,
    author = {Cruz-Martinez, Juan M. and Fieg, Max and Giani, Tommaso and Krack, Peter and M{\"a}kel{\"a}, Toni and Rabemananjara, Tanjona R. and Rojo, Juan},
    title = "{The LHC as a Neutrino-Ion Collider}",
    eprint = "2309.09581",
    archivePrefix = "arXiv",
    primaryClass = "hep-ph",
    reportNumber = "Nikhef-2023-009, CERN-TH-2023-165",
    doi = "10.1140/epjc/s10052-024-12665-1",
    journal = "Eur. Phys. J. C",
    volume = "84",
    number = "4",
    pages = "369",
    year = "2024"
}

@techreport{Albanese:2948477,
    author        = "Albanese, R and others",
      collaboration = "SHiP",
      title         = "{BDF/SHiP Annual Report 2025}",
      institution   = "CERN",
      reportNumber  = "CERN-SPSC-2025-039, SPSC-SR-370",
      address       = "Geneva",
      year          = "2025",
      url           = "https://cds.cern.ch/record/2948477",
}

\onecolumn
\begin{center}
\textbf{The SND@LHC Collaboration}
\vspace{0.25cm}
\break
\author{D.~Abbaneo$^{9}$\orcidlink{0000-0001-9416-1742}},
\author{S.~Ahmad$^{42}$\orcidlink{0000-0001-8236-6134}},
\author{R.~Albanese$^{1,2}$\orcidlink{0000-0003-4586-8068}},
\author{A.~Alexandrov$^{1}$\orcidlink{0000-0002-1813-1485}},
\author{F.~Alicante$^{1,2}$\orcidlink{0009-0003-3240-830X}},
\author{F.~Aloschi$^{1,2}$\orcidlink{0000-0002-2501-7525}},
\author{K.~Androsov$^{6}$\orcidlink{0000-0003-2694-6542}},
\author{A.~Anokhina$^{3}$\orcidlink{0000-0002-4654-4535}},
\author{L.G.~Arellano$^{1,2}$\orcidlink{0000-0002-1093-1824}},
\author{C.~Asawatangtrakuldee$^{38}$\orcidlink{0000-0003-2234-7219}},
% ...
\end{center}

\begin{flushleft}
\begin{footnotesize}
$^{1}$Sezione INFN di Napoli, Napoli, 80126, Italy\linebreak
$^{2}$Universit\`{a} di Napoli ``Federico II'', Napoli, 80126, Italy\linebreak
$^{3}$Affiliated with an institute formerly covered by a cooperation agreement with CERN\linebreak
% ...
$^{6}$Institute of Physics, EPFL, Lausanne, 1015, Switzerland\linebreak
% ...
$^{9}$European Organization for Nuclear Research (CERN), Geneva, 1211, Switzerland\linebreak
% ...
$^{38}$Department of Physics, Faculty of Science, Chulalongkorn University, Bangkok, 10330, Thailand\linebreak
% ...
$^{42}$Affiliated with Pakistan Institute of Nuclear Science and Technology (PINSTECH), Nilore, 45650, Islamabad, Pakistan
% ...
\end{footnotesize}
\end{flushleft}
\end{document}